\documentclass{article}

\usepackage[preprint]{neurips_2022}

\usepackage[utf8]{inputenc} 
\usepackage[T1]{fontenc}    
\usepackage{hyperref}       
\usepackage{url}            
\usepackage{booktabs}       
\usepackage{amsfonts}       
\usepackage{nicefrac}       
\usepackage{microtype}      
\usepackage{xcolor}         
\usepackage{graphicx}
\usepackage{kotex}
\usepackage{amsmath}
\usepackage{amssymb}
\usepackage{mathtools}
\usepackage{amsthm}
\usepackage{multirow}

\newcommand{\norm}[1]{\left\lVert#1\right\rVert}
\newcommand{\shortcite}[1]{[\citenum{#1}]}
\newcommand{\namecite}[1]{\citeauthor{#1} [\citenum{#1}]}
\setcitestyle{citesep={,}}

\title{Guided-TTS 2: A Diffusion Model for High-quality Adaptive Text-to-Speech with Untranscribed Data}

\author{%
  Sungwon Kim\thanks{Equal contribution.} \\
  Data Science \& AI Lab.\\
  Seoul National University \\
  \texttt{ksw0306@snu.ac.kr} \\
  \And
  Heeseung Kim\footnotemark[1]\\
  Data Science \& AI Lab.\\
  Seoul National University \\
  \texttt{gmltmd789@snu.ac.kr} \\
  \And
  Sungroh Yoon\thanks{Corresponding author}\\
  Data Science \& AI Lab.\\
  Seoul National University \\
  \texttt{sryoon@snu.ac.kr} \\
}

\begin{document}

\maketitle

\begin{abstract}
  We propose Guided-TTS 2, a diffusion-based generative model for high-quality adaptive TTS using untranscribed data. Guided-TTS 2 combines a speaker-conditional diffusion model with a speaker-dependent phoneme classifier for adaptive text-to-speech. We train the speaker-conditional diffusion model on large-scale untranscribed datasets for a classifier-free guidance method and further fine-tune the diffusion model on the reference speech of the target speaker for adaptation, which only takes 40 seconds. We demonstrate that Guided-TTS 2 shows comparable performance to high-quality single-speaker TTS baselines in terms of speech quality and speaker similarity with only a ten-second untranscribed data. We further show that Guided-TTS 2 outperforms adaptive TTS baselines on multi-speaker datasets even with a zero-shot adaptation setting. Guided-TTS 2 can adapt to a wide range of voices only using untranscribed speech, which enables adaptive TTS with the voice of non-human characters such as Gollum in \textit{"The Lord of the Rings"}.
\end{abstract}

\section{Introduction}
Neural speech synthesis models have shown to produce natural samples \shortcite{shen2018natural, van2016wavenet, wang17n_interspeech}. Recent two-stage TTS models as well as end-to-end TTS models generate high-quality speech in real time. \shortcite{wavegrad2, donahue2021endtoend, kim2020glow, kim2021conditional, kimflowavenet, prenger2019waveglow, ren2019fastspeech, ren2021fastspeech}. However, these high-quality TTS models assume that sufficient amounts of large-scale data for the target speaker are given, which is a large constraint. Adaptive TTS models aim to generate high-quality speech for the target speaker given a limited amount of reference data. In general, the more reference data is given, the better adaptive TTS models adapt to the target speaker, and one of the main challenges is to reduce the amount of reference data while preserving adaptation quality and naturalness.

Most adaptive TTS models learn to generate speech from various speakers using a multi-speaker dataset and adapt to new speakers through various methods. There have been many zero-shot adaptive TTS models using a pre-trained speaker verification model to extract the speaker embedding from a short reference speech or their own speaker encoder to extract more generalizable speaker representation \shortcite{casanova21b_interspeech, 2021arXiv211202418C, cooper2020zero, wu2022adaspeech}. Fine-tuning based adaptive TTS models have also been proposed to update part of pre-trained multi-speaker TTS models using a small amount of reference speech \shortcite{2021arXiv211202418C, chen2021adaspeech, chen2018sample, yan2021adaspeech}. \namecite{pmlr-v139-min21b} propose meta-learning based adaptation using short-duration reference speech. However, when the amount of the reference speech is only about 10-seconds-long, existing adaptive TTS models show poor sample quality and speaker similarity compared to single-speaker high-quality TTS models trained on a single-speaker TTS dataset, such as LJSpeech.

Recent diffusion-based generative models show impressive results in class-conditional and text-conditional image generation tasks via diffusion guidance methods \shortcite{dhariwal2021diffusion, DBLP:journals/corr/abs-2112-10741,  https://doi.org/10.48550/arxiv.2204.06125}. In speech synthesis, diffusion-based models generate high-quality audio samples while providing trade-off between sample quality and inference speed \shortcite{WaveGAN, lee2022priorgrad, jeong21_interspeech, DiffWave, Grad-TTS}. Using the modified classifier guidance method, Guided-TTS \shortcite{DBLP:journals/corr/abs-2111-11755} enables high-quality TTS using untranscribed data of the target speaker. Despite the success of diffusion models for high-quality text-to-speech synthesis, the potentials of diffusion models for adaptive TTS have not yet been explored. 

In this work, we present Guided-TTS 2, which introduces a diffusion-based generative model for adaptive TTS with a small amount of untranscribed data. We train a speaker-conditional DDPM on large-scale untranscribed datasets to model the distribution of speech for various speakers. Similar to Guided-TTS \shortcite{DBLP:journals/corr/abs-2111-11755}, Guided-TTS 2 performs text-to-speech synthesis by guiding the sampling process of the speaker-conditional DDPM with the pre-trained phoneme classifier. We adapt our pre-trained diffusion model to the target speaker by using the classifier-free guidance method and fine-tuning directly on the reference speech data. Although the fine-tuning approach requires additional training time for adaptation, Guided-TTS 2 significantly reduces the fine-tuning cost to less than a minute on top of our speaker-conditional diffusion model's adaptability.

\begin{figure*}[t]
    \centering
    \includegraphics[width=0.9\linewidth]{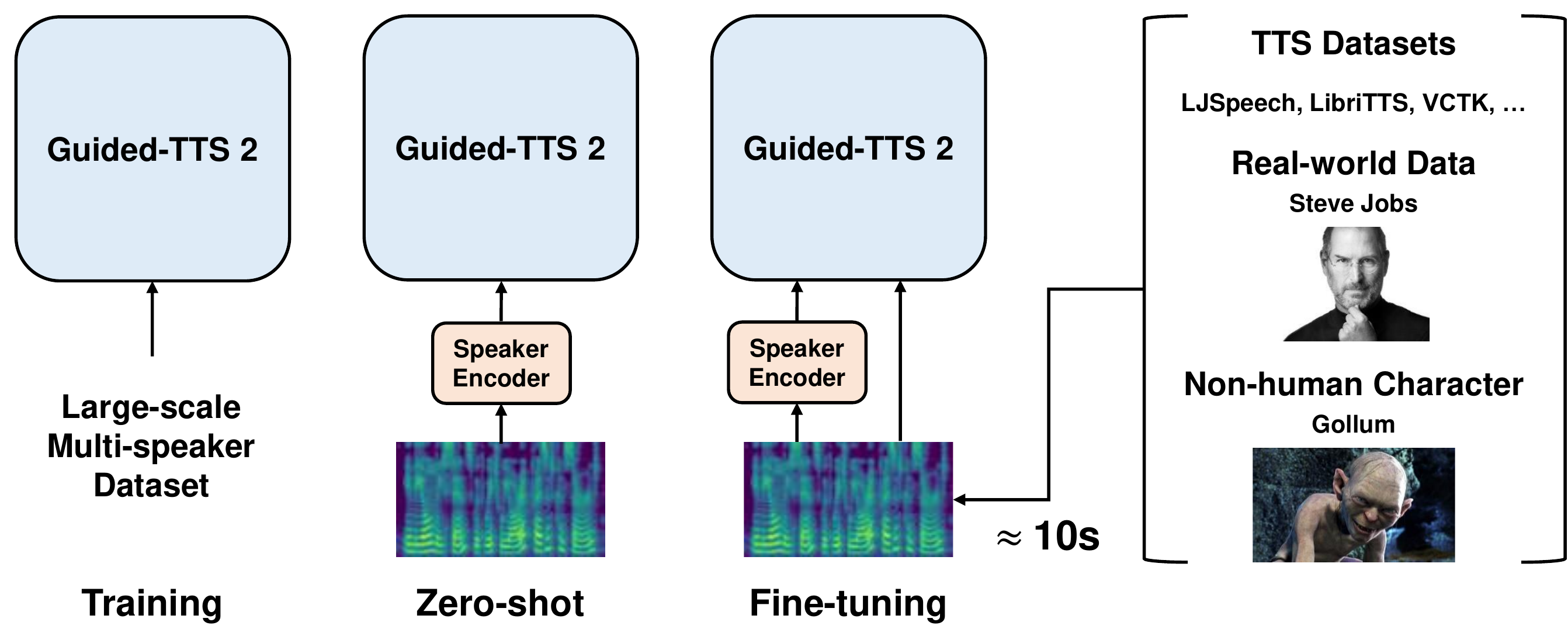}
    \caption{The overview of Guided-TTS 2.}
    \label{fig1}
    \vskip -0.2in
\end{figure*}

We demonstrate that Guided-TTS 2 fine-tuned with a ten-second reference speech of LJSpeech obtains comparable sample quality and speaker similarity to the single-speaker high-quality TTS baselines trained on 24-hour LJSpeech. Furthermore, Guided-TTS 2 outperforms the adaptive TTS baselines, YourTTS and Meta-StyleSpeech, for the LibriTTS and VCTK datasets. As our speaker-conditional DDPM adapts without transcript, Guided-TTS 2 enables adaptive TTS with audio clips that are difficult to transcribe, such as the voice of Gollum in \textit{"The Lord of The Rings"}. We highly encourage readers to listen to various samples of Guided-TTS 2 shown in Fig. \ref{fig1} on our demo page.\footnote{Demo: \href{https://bit.ly/3wUBeIm}{https://bit.ly/3wUBeIm}}
\section{Background}
\subsection{Score-based Generative Model}
\namecite{song2021scorebased} introduce a unified approach of denoising diffusion probabilistic model (DDPM) \shortcite{DDPM, pmlr-v37-sohl-dickstein15} and score matching with Langevin dynamics (SMLD) \shortcite{NEURIPS2019_3001ef25, NEURIPS2020_92c3b916}. We follow the formulation of Grad-TTS \shortcite{Grad-TTS}, which applies \namecite{song2021scorebased} to TTS. \namecite{Grad-TTS} define the diffusion process, which can convert any data distribution to standard normal distribution
\begin{equation}
    \label{forward diffusion}
    dX_t = -\frac{1}{2}X_t\beta_tdt+\sqrt{\beta_t}dW_t,\quad t\in[0,T],
\end{equation}
where $\beta_t$ is the pre-defined noise schedule $\beta_t=\beta_0+(\beta_T-\beta_0)t$ and $W_t$ is the Wiener process. \namecite{Anderson1982-ny} proves that the reverse process, which follows the reverse trajectory of the diffusion process, can be formulated in SDE. \namecite{Grad-TTS} use the following equation, which is a discretized version of the reverse SDE, during sampling to generate data $X_0$ from standard Gaussian noise $X_T$
\begin{equation}
    \label{discretized reverse diffusion}
    X_{t-\frac{1}{N}} = X_t + \frac{\beta_t}{N}(\frac{1}{2}X_t + s(X_t)) + \sqrt{\frac{\beta_t}{N}}z_t,\quad s(X_t)=\nabla_{X_t}\log{p_t(X_t)},
\end{equation}
where $N$ is the number of steps in the discretized reverse process, and $z_t$ is the standard Gaussian noise. As \namecite{Grad-TTS} set $T$ to 1, the size of one step is $\frac{1}{N}$, and $t \in \{\frac{1}{N}, \frac{2}{N}, ..., 1\}$.

As can be seen in Eq. \ref{discretized reverse diffusion}, $\nabla_{X_t}\log{p_t(X_t)}$, called the score $s(X_t)$, is required for sampling procedure. To predict the score, data $X_0$ is first corrupted into $X_t$ at an arbitrary timestep $t$ using $\epsilon_t\sim N(0,I)$ and Eq. \ref{forward diffusion}. Afterwards, a neural network $s_\theta$ is trained using the following loss to estimate the score
\begin{equation}
    \label{loss}
    L(\theta) = \mathbb{E}_{t,X_0,\epsilon_t}\big[\norm{s_\theta(X_t,t)+\lambda(t)^{-1}\epsilon_t}_2^2\big],
\end{equation}
where $\lambda(t)=I-{\rm e}^{-\int_0^t\beta_{s}ds}$. By minimizing this loss, the model $s_\theta$ is trained to compute the score, which is essential for generating data from noise.

\subsection{Diffusion Guidance Methods}
Recently, classifier guidance \shortcite{dhariwal2021diffusion, song2021scorebased} and classifier-free guidance \shortcite{ho2021classifierfree} are proposed for guiding the diffusion model. The discretized reverse process for each method can be obtained by replacing the unconditional score $s(X_t)$ in Eq. \ref{discretized reverse diffusion} with the modified conditional score $\hat{s}(X_t|y)$, which is defined differently depending on the approach.

\textbf{Classifier Guidance} \namecite{song2021scorebased} show that it is possible to generate conditional samples using unconditional diffusion model by leveraging the separately trained classifier. Given the unconditional score and the gradient of the log-density of the classifier $\nabla_{X_t}\log{p_t(y|X_t)}$, the conditional score $s(X_t|y)$ can be estimated by adding the unconditional score and the classifier gradient. Conditional samples can then be generated by setting the modified conditional score equal to the conditional score as in Eq. \ref{classifier uncon}.
\begin{equation}
    \label{classifier uncon}
    \hat{s}(X_t|y) = s(X_t|y) = \nabla_{X_t}\log{p_t(X_t|y)}=s(X_t)+\nabla_{X_t}\log{p_t(y|X_t)}.
\end{equation}
Classifier guidance can also be utilized to improve conditional diffusion models. \namecite{dhariwal2021diffusion} provide an additional classifier gradient to the conditional score, and the modified conditional score accordingly is as follows:
\begin{equation}
    \label{classifier con}
    \hat{s}(X_t|y) = s(X_t|y)+\gamma\cdot\nabla_{X_t}\log{p_t(y|X_t)}.
\end{equation}
\namecite{dhariwal2021diffusion} show that sample quality can be improved at the expense of diversity if more classifier gradient is provided by increasing the gradient scale $\gamma$. By adjusting the gradient scale for the optimal value, they achieve state-of-the-art performance in conditional image generation.

\textbf{Classifier-free Guidance} Unlike classifier guidance, classifier-free guidance \shortcite{ho2021classifierfree} does not require a classifier to guide a diffusion model toward a target condition. By learning the conditional and unconditional scores with a single model, we can calculate the classifier gradient by subtracting the unconditional score from the conditional score. During training, the condition $y$ is sometimes replaced with the null label $\phi$, making it possible to jointly train unconditional and conditional models. The modified conditional score for classifier-free guidance is as follows:
\begin{equation}
    \label{classifier-free}
    \hat{s}(X_t|y) = s(X_t|y) +\gamma\cdot(s(X_t|y)-s(X_t|\phi)).
\end{equation}
According to GLIDE \shortcite{DBLP:journals/corr/abs-2112-10741}, classifier-free guidance has two main advantages. The first advantage is that classifier-free guidance enables guiding the diffusion model with only a single model. Second, the diffusion model can be guided even toward conditions that are difficult to predict using a classifier such as text. In GLIDE, diffusion model with classifier-free guidance achieves better performance than with CLIP-based \shortcite{pmlr-v139-radford21a} classifier guidance in text-to-image generation. Similarly, DALL$\cdot$E 2 \shortcite{https://doi.org/10.48550/arxiv.2204.06125} shows powerful performance in text-to-image generation task using CLIP and classifier-free guidance.

\subsection{Guided-TTS}
Guided-TTS \shortcite{DBLP:journals/corr/abs-2111-11755} is a model that constructs high-quality TTS without any transcript of the target speaker. \namecite{DBLP:journals/corr/abs-2111-11755} train an unconditional DDPM using the untranscribed speech and guide it using a phoneme classifier and duration predictor trained on a large-scale multi-speaker ASR dataset. \namecite{DBLP:journals/corr/abs-2111-11755} provide speaker embedding extracted from the speaker encoder to the phoneme classifier and duration predictor, making both modules speaker-dependent. These speaker-dependent modules are generalizable to various unconditional models of unseen speakers, enabling a single phoneme classifier and duration predictor to guide multiple unconditional DDPMs.

\namecite{DBLP:journals/corr/abs-2111-11755} observe that applying the existing classifier guidance methods \shortcite{song2021scorebased, dhariwal2021diffusion} to text-to-speech cannot guarantee accurate pronunciation regardless of the gradient scale. In order to generate speech that faithfully reflects the given text, \namecite{DBLP:journals/corr/abs-2111-11755} propose norm-based guidance, a classifier guidance method of scaling the norm of the classifier gradient accordingly to the norm of the score, which is as follows:
\begin{equation}
    \label{norm-based guidance}
    \hat{s}(X_t|y) = s(X_t)+\gamma\cdot\frac{\norm{s(X_t)}}{\norm{\nabla_{X_t}\log{p_t(y|X_t)}}}\cdot\nabla_{X_t}\log{p_t(y|X_t)}.
\end{equation}
With the norm-based guidance, \namecite{DBLP:journals/corr/abs-2111-11755} show that Guided-TTS is comparable to other TTS models in terms of audio quality and pronunciation accuracy, showing the possibility of building a high-quality TTS model using untranscribed speech.
\section{Guided-TTS 2}
\begin{figure*}[h]
    \centering
    \includegraphics[width=0.9\linewidth]{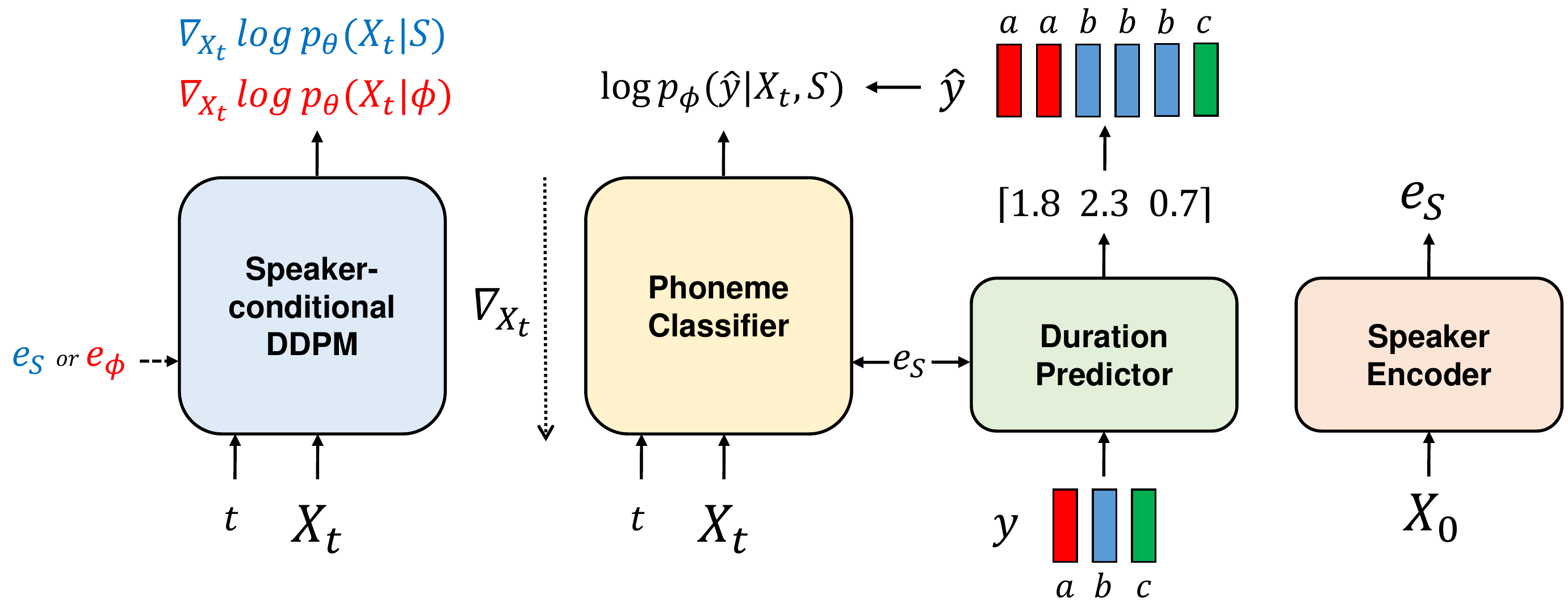}
    \caption{The overall components of Guided-TTS 2.}
    \label{fig2}
    \vskip -0.1in
\end{figure*}
In this section, we present Guided-TTS 2, which aims to adapt new speakers with a small amount of untranscribed data. Extending Guided-TTS to adaptive TTS, we train a speaker-conditional DDPM on multi-speaker untranscribed datasets and guide the diffusion model with the phoneme classifier using classifier guidance. Based on the pronunciation accuracy of the norm-based guidance in Guided-TTS, we focus on the speaker-conditional DDPM for estimating the conditional score for new speakers.

We pre-train our speaker-conditional DDPM on large-scale multi-speaker datasets for speaker classifier-free guidance. For better adaptation to new speakers, we further fine-tune the diffusion model using the untranscribed reference speech of the target speaker. Since Guided-TTS 2 adapts only with the untranscribed speech, Guided-TTS 2 can enable adaptive TTS using challenging reference speeches that may contain auditory contents untranscribable in the desired language.

In Section \ref{sec:3.1}, we introduce the speaker-conditional DDPM and speaker classifier-free guidance. We present the details of fine-tuning for adaptation to new speakers in Section \ref{sec:3.2}, and describe components of Guided-TTS 2 and sampling process for adaptive TTS in Section \ref{sec:3.3}. 

\subsection{Speaker-conditional Diffusion Model}\label{sec:3.1}
Our speaker-conditional DDPM estimates speaker-conditional score of speech for various speakers without any transcript. For zero-shot adaptation ability to target speakers, we use a pre-trained speaker verification model as our speaker encoder $E_S$ like \namecite{Verification}, which encodes speaker information of untranscribed data.

Given a mel-spectrogram $X=X_0$ in a large-scale unlabeled dataset, let us define a forward diffusion process of $X_t, t \in [0,1]$ as in Eq. \ref{forward diffusion}. Our speaker-conditional DDPM learns to estimate conditional score $\nabla_{X_t}\log P_\theta(X_t|S)$ for speaker $S$ that is required for the reverse denoising process. We use the same U-Net architecture \shortcite{ronneberger2015u} of unconditional DDPM in Guided-TTS \shortcite{DBLP:journals/corr/abs-2111-11755} as our speaker-conditional DDPM $s_\theta$. For conditioning speaker information, we extract a speaker embedding $e_S=E_S(X)\in R^d$ from the mel-spectrogram $X$, and concatenate speaker embedding $e_S$ with timestep $t$ for speaker conditioning. The training objective of speaker-conditional DDPM $s_\theta$ is as follows: 
\begin{equation}
\label{speaker-condtional DDPM}
    L(\theta) = \mathbb{E}_{t,X_0,\epsilon_{t}}[\|{s_\theta (X_t|S)+\lambda(t)^{-1}\epsilon_t}\|_2^2].
\end{equation}
We further introduce a classifier-free guidance \shortcite{ho2021classifierfree} to augment the speaker condition from the speaker embedding. For speaker classifier-free guidance, as in GLIDE \shortcite{DBLP:journals/corr/abs-2112-10741}, we introduce the null embedding $e_\phi \in R^d$ and additionally train the speaker-conditional DDPM to estimate both the conditional and the unconditional score. At every iteration, we replace the speaker embedding $e_S$ of each data with a null embedding $e_\phi$ with a $50\%$ chance to output the unconditional score $s_\theta(X_t|\phi)$. As our speaker embedding is set to have a unit norm, we define the null embedding using additional trainable parameter $w\in R^d$ as $e_{\phi}=\frac{w}{\|w\|}$ for consistency.

For zero-shot adaptation, we extract speaker embedding $e_{\hat{S}}$ from the reference untranscribed data of the target speaker $\hat{S}$ and provide speaker embedding $e_{\hat{S}}$ to the diffusion model for adaptation. We augment the speaker condition of the target speaker by setting the classifier-free scale $\gamma_S$ to be larger than $0$ and compute the modified conditional score $\hat{s}_\theta(X_t|\hat{S})$ as in Eq. \ref{classifier_free guidance inference}. 
\begin{equation}
\label{classifier_free guidance inference}
    \hat{s}_\theta(X_t|\hat{S}) = s_\theta(X_t|\hat{S}) + \gamma_S * (s_\theta(X_t|\hat{S}) - s_\theta(X_t|\phi))
\end{equation}
\subsection{Fine-tuning}\label{sec:3.2}
The zero-shot adaptation method described above relies on the speaker embedding extracted from the reference data to adapt to the target speaker. To better adapt to the target speaker, we directly fine-tune the pre-trained speaker-conditional DDPM on the reference data to estimate the speaker-conditional score of the target speaker. While most existing adaptive TTS models require more than a minute of reference data for fine-tuning, we found that speaker-conditional DDPM trained on a large-scale multi-speaker dataset can adapt surprisingly well to the target speaker even only with a ten-second reference data. However, if speaker-conditional diffusion overfits to the single reference data, the diffusion model loses the pronunciation ability learned from the multi-speaker dataset and is difficult to use for TTS synthesis. Therefore, we fine-tune the diffusion model for a small number of iterations with a small learning rate so as not to overfit to the single reference data.

Given the reference untranscribed data for the target speaker $\hat{S}$, we first extract the speaker embedding $e_{\hat{S}}$ from the reference data. Using the reference data, we optimize the diffusion model through the Adam optimizer with a learning rate of $2\cdot10^{-5}$, which is lower than the learning rate $10^{-4}$ used for pre-training. We found that the statistics of the pre-trained optimizer degrade the adaptation performance. Thus, we reuse only the pre-trained weights of the diffusion model and initialize the optimizer for fine-tuning.
We fine-tune the speaker-conditional DDPM to estimate only the conditional score for the target speaker as in Eq. \ref{speaker-condtional DDPM}. 

We update the diffusion model by 500 iterations, to the extent that the model adapts well to various target speakers while preserving pronunciation ability. The following training time is about 40 seconds on a NVIDIA RTX 8000 GPU, which is small enough to build a TTS model for the target speaker. 

\subsection{Adaptive Text-to-Speech}\label{sec:3.3}
 In this section, we explain each module of Guided-TTS 2 and describe how Guided-TTS 2 performs adaptive TTS using reference speech. Similar to \namecite{DBLP:journals/corr/abs-2111-11755}, Guided-TTS 2 consists of a speaker-conditional DDPM, a phoneme classifier, a duration predictor, and a speaker encoder, as shown in Fig. \ref{fig2}. The architecture of each module is detailed in Section \ref{app::architectures}. The phoneme classifier predicts the phoneme sequence frame-wise from the noisy mel-spectrogram $X_t$, and the duration predictor outputs the duration of each phoneme. All modules use speaker embedding as a condition for generalization to new speakers. In general, we use a fine-tuned diffusion model in Section \ref{sec:3.2} for adaptation except for zero-shot adaptation, which we use the pre-trained speaker-conditional DDPM described in Section \ref{sec:3.1}.

Let us assume that the reference data for the target speaker $\hat{S}$ and the phoneme sequence $y$ to generate are given. We first predict the duration of each phoneme with the duration predictor, and expand the phoneme sequence $y$ to frame-level phoneme sequence $\hat{y}$ with the predicted duration. Then, we sample the noise $X_1$ having the same length as the frame-level phoneme sequence from the standard Gaussian distribution for the reverse process of DDPM. In order to perform adaptive TTS, we replace $s(X_t)$ in Eq. \ref{discretized reverse diffusion} with the modified conditional score $\hat{s}_\theta(X_t|\hat{y}, \hat{S})$ for the target speaker $\hat{S}$ and the frame-level phoneme sequence $\hat{y}$ during sampling. 

To compute the modified conditional score $\hat{s}_\theta(X_t|\hat{y}, \hat{S})$, we combine the speaker classifier-free guidance in Eq. \ref{classifier_free guidance inference} and the norm-based guidance for text in Eq. \ref{norm-based guidance}. At each timestep in the reverse sampling process, we obtain $\hat{s}_\theta(X_t|\hat{S})$ from the speaker-conditional DDPM through the speaker classifier-free guidance, and compute the classifier gradient $\nabla_{X_t}\log{p_t(\hat{y}|X_t, \hat{S})}$ from the pre-trained phoneme classifier. With $\hat{s}_\theta(X_t|\hat{S})$ and $\nabla_{X_t}\log{p_t(\hat{y}|X_t, \hat{S})}$, we compute the modified conditional score for the target speaker and the phoneme sequence using the norm-based guidance as follows:
\begin{equation}
\label{Guided-TTS 2 guidance}
    \hat{s}_\theta(X_t|\hat{y}, \hat{S}) = \hat{s}_\theta(X_t|\hat{S}) + \gamma_T\cdot\frac{\norm{\hat{s}_\theta(X_t|\hat{S})}}{\norm{\nabla_{X_t}\log{p_t(\hat{y}|X_t, \hat{S})}}}\cdot\nabla_{X_t}\log{p_t(\hat{y}|X_t, \hat{S})}, 
\end{equation}
where $\gamma_T$ is the text gradient scale for the norm-based guidance method. 
\section{Experiments}
\label{experiments}
\textbf{Datasets} The speaker-dependent phoneme classifier and duration predictor used in Guided-TTS 2 are trained on LibriSpeech \shortcite{7178964}, a multi-speaker ASR dataset of approximately 1,000 hours. We train the speaker encoder on VoxCeleb2 \shortcite{Voxceleb2}, a dataset with over 1M utterances of 6,112 speakers.

We use LibriTTS \shortcite{zen19_interspeech} and Libri-Light \shortcite{librilight} to train the speaker-conditional diffusion model. In addition, to compare the performance according to the amount of data used for training, we also train the diffusion model only on LibriTTS. LibriTTS is a large-scale multi-speaker TTS dataset consisting of approximately 585 hours of labeled speech spoken by 2,456 speakers. Libri-Light is a dataset composed of 60K hours of unlabeled speech and a small amount of labeled speech (10 hours, 1 hour, 10 minutes) obtained from audiobooks. Among Libri-Light, we use \texttt{unlab-6k} and \texttt{unlab-600} in the unlabeled training set for training, approximately 6,300 hours of unlabeled speech spoken by 2,231 speakers. Unlike LibriTTS, the length of each sample in Libri-Light is too long to be used for training. We thus randomly cut each sample into 5-seconds-long and use it to train the diffusion model.

To evaluate adaptation performance, we only use unseen speakers for reference speakers during training for comparison. To compare with single-speaker TTS models, we use LJSpeech \shortcite{ljspeech17}, a female speaker dataset consisting of 13,100 audio clips. We randomly sample 10 audios among LJSpeech with a length of 9 to 11 seconds and use them as reference data for Guided-TTS 2. To compare with the adaptive TTS model, we use the LibriTTS and VCTK \shortcite{Yamagishi2019CSTRVC}, a paired dataset of approximately 44 hours uttered by 110 speakers. We select the same speakers and the same reference audios as YourTTS \shortcite{2021arXiv211202418C}. That is, we use reference data of 11 speakers, 7 women and 4 men from each accent, from VCTK, and reference data of 10 speakers from subset \texttt{test-clean} of LibriTTS. In addition, we perform TTS with Guided-TTS 2 using real-world reference audio. We extract 10-second audio clips of several speakers from Youtube and use them as reference clips. As we do not have licenses for real-world scenarios to re-distribute, we instead provide Youtube links for each reference audio.

\textbf{Training and Fine-tuning Details} We train the model using the same architectures and hyperparameters as Guided-TTS except for the speaker-conditional diffusion model. As in Guided-TTS, we convert text into International Phonetic Alphabet (IPA) phoneme sequences using open-source software \shortcite{phonemizer20} and convert the speech into mel-spectrogram with the same hyperparameters as Glow-TTS \shortcite{kim2020glow}. We extract the alignment of speech and transcript required for training the phoneme classifier and duration predictor using Montreal Forced Aligner (MFA) \shortcite{MFA}. The phoneme classifier, duration predictor, and speaker encoder are all trained in the same way as Guided-TTS. 

During diffusion model pre-training, we first train speaker-conditional DDPM using LibriTTS for 1M iterations with batch size 256 on 4 NVIDIA RTX 8000 GPUs. We additionally train the pre-trained DDPM for classifier-free guidance in two different ways: (i) using LibriTTS only and (ii) using LibriTTS and Libri-Light. The former case is expressed as Guided-TTS 2 (LT) and the latter case is expressed and Guided-TTS 2 (LT+LL). For both cases, we train the models for 100K iterations. 

\textbf{Evaluation} We use Grad-TTS and Guided-TTS as single-speaker TTS baselines. For Grad-TTS, we use the official implementation and the pre-trained model\footnote{Grad-TTS: \href{https://bit.ly/3NaSL4n}{https://bit.ly/3NaSL4n}}. We set the temperature and the number of reverse steps of Grad-TTS, Guided-TTS, and Guided-TTS 2 to $\tau=1.5$ and $N=50$. For Guided-TTS 2, we set text gradient scale $\gamma_T$ to $0.3$ and speaker gradient scale $\gamma_S$ to $1.0$. For the vocoder, we use the official implementation and pre-trained model of universal HiFi-GAN.\footnote{HiFi-GAN: \href{https://bit.ly/3w9OHf1}{https://bit.ly/3w9OHf1}} We set the sampling rate of ground truth audio and the samples from single-speaker TTS models to 22,050Hz.

For comparison to adaptive TTS models, we use the official implementation and pre-trained model of zero-shot adaptive TTS models, YourTTS and Meta-StyleSpeech \shortcite{pmlr-v139-min21b}.\footnote{YourTTS: \href{https://bit.ly/3wbC6I1}{https://bit.ly/3wbC6I1}, Meta-StyleSpeech: \href{https://bit.ly/3FJMSsg}{https: //bit.ly/3FJMSsg}} Unlike YourTTS, which is an end-to-end TTS model, Meta-StyleSpeech requires a vocoder. Therefore, we train universal HiFi-GAN that matches the frontend of Meta-StyleSpeech using an official implementation of HiFi-GAN. Since YourTTS and Meta-StyleSpeech generate speech with a sampling rate of 16kHz, for a fair comparison, we downsample the speech generated by Guided-TTS 2 to 16kHz and then compare it with baselines.

To show the pronunciation accuracy of our model, we measure the character error rate (CER). The CTC-based conformer of the NEMO toolkit \shortcite{kuchaiev2019nemo} trained in 7K hours of speech is used for measuring the CER. To compare the performance of models according to the speaker classifier-free gradient scale $\gamma_S$, we measure the CER and speaker encoder cosine similarity (SECS). For the SECS measurement, as in YourTTS, we use speaker embedding extracted from the speaker encoder of the Resemblyzer \shortcite{resemblyzer} package. We generate each sentence of the test set 5 times, and use the averaged metrics values for each model.
\section{Results}
\subsection{Comparison to Single Speaker TTS Models}\label{results::single}
To compare the audio quality and speaker similarity with reference audio, we measure the 5-scale mean opinion score (MOS) and the 5-scale speaker similarity mean opinion score (SMOS) using Amazon Mechanical Turk for LJSpeech. In addition, we check the pronunciation accuracy of each model through CER. We randomly select 50 samples from LJSpeech's test set for evaluation. We use 10 pre-selected reference samples for measuring SMOS. For Guided-TTS 2, we generate 5 sentences per reference audio.

\begin{table}[h]
\caption{Mean opinion score (MOS) and speaker similarity mean opinion score (SMOS) with 95$\%$ confidence intervals of single-speaker TTS models and Guided-TTS 2 for LJSpeech. The diffusion model of Guided-TTS 2 (LT+LL) is trained on LibriTTS and Libri-Light, whereas Guided-TTS 2 (LT) is trained on LibriTTS.}
\label{mos_LJSpeech}
\vskip -0.1in
\begin{center}
\begin{tabular}{lccc}
\toprule
\multicolumn{1}{c}{\bf Method} &\multicolumn{1}{c}{\bf 5-scale MOS} & \multicolumn{1}{c}{\bf CER(\%)} &\multicolumn{1}{c}{\bf 5-scale SMOS}
\\ \hline 
Ground Truth    & 4.45$\pm$0.05 & 0.64& 3.85$\pm$0.08\\
Mel + HiFi-GAN (\cite{kong2020hifi})& 4.24$\pm$0.08 & 0.86 & 3.80$\pm$0.08\\
Grad-TTS (\cite{Grad-TTS})    &  4.22$\pm$0.08 & 0.98& 3.67$\pm$0.09\\
Guided-TTS (\cite{DBLP:journals/corr/abs-2111-11755})&  4.17$\pm$0.09 & 1.23& 3.63$\pm$0.09\\
\hline
Guided-TTS 2 (LT+LL) &  4.21$\pm$0.09 & 1.12& 3.69$\pm$0.09\\
Guided-TTS 2 (LT+LL zero-shot) & 4.23$\pm$0.09 & 0.89& 3.51$\pm$0.08\\
Guided-TTS 2 (LT) &  4.22$\pm$0.09 & 1.16& 3.74$\pm$0.09\\
Guided-TTS 2 (LT zero-shot) & 4.16$\pm$0.09 & 1.03& 3.47$\pm$0.09\\
\bottomrule
\end{tabular}
\end{center}
\vskip -0.1in
\end{table}
In Table \ref{mos_LJSpeech}, we compare Guided-TTS 2 with the high-quality single speaker TTS models, Grad-TTS and Guided-TTS. Unlike the baselines, which fully utilize LJSpeech for training, Guided-TTS 2 uses a 10-second randomly chosen reference clip from LJSpeech. Through MOS and CER, we confirm that both Guided-TTS 2 and Guided-TTS 2 (zero-shot) generate high-quality samples with accurate pronunciation regardless of fine-tuning. In particular, Guided-TTS 2 with fine-tuning shows comparable performance in all metrics, including the metric for speaker similarity, to the existing TTS model. These experimental results demonstrate that we can build a high-quality TTS model, even on par with the high-quality single-speaker model, with only 10-second untranscribed data.

We additionally show that existing adaptive models are inferior to single-speaker TTS models while Guided-TTS 2 achieves competitive performance, which is very challenging for the case of 10-seconds long reference audio. We compare the performance of the single speaker TTS models, the adaptive TTS models, and Guided-TTS 2 for LJSpeech in Section \ref{app::LJ_16kHz}. While the existing adaptive TTS models are far behind in sample quality and speaker similarity compared to the single-speaker TTS models, Guided-TTS 2 showed comparable performance to single-speaker TTS despite being an adaptive TTS model. Samples of all models are on the demo page.

\textbf{Amount of Data for Training} We do not need any transcript to train the speaker-conditional diffusion model of Guided-TTS 2. Therefore, we can use unlabeled datasets such as Libri-Light, which the existing adaptive TTS models cannot utilize. In order to check the performance according to the amount of training data, we compare Guided-TTS 2 (LT) with Guided-TTS 2 (LT+LL), where the former uses LibriTTS for additional training for classifier-free guidance and the latter uses LibriTTS and Libri-Light together. As shown in Table \ref{mos_LJSpeech}, we observe that LibriTTS is sufficient for fine-tuning-based adaptation. We also find that using more untranscribed data is beneficial for our model in zero-shot performance. From this section, we use Guided-TTS 2 (LT+LL) for evaluation, which uses both LibriTTS and Libri-Light for training.
\subsection{Comparison to Adaptive TTS Models}\label{results::adaptive}
In this section, we compare our model with adaptive TTS models given a variety of reference audios from diverse datasets by measuring MOS, SMOS, are CER. We select 55 samples and 50 samples from the test set of VCTK and LibriTTS, respectively. We use the same reference data as YourTTS, 11 reference audios for VCTK, each from 11 respective reference speakers, and 10 reference audios for LibriTTS, each from 10 respective reference speakers. We generate 5 sentences per reference audio.
\begin{table}[h]
\caption{Mean opinion score (MOS) and speaker similarity mean opinion score (SMOS) with 95$\%$ confidence intervals of adaptive TTS models and Guided-TTS 2 for LibriTTS and VCTK.}
\vskip -0.2in
\label{mos_multi}
\begin{center}
\begin{tabular}{clccc}
\toprule
\multicolumn{1}{c}{\bf Dataset}&\multicolumn{1}{c}{\bf Method} &\multicolumn{1}{c}{\bf 5-scale MOS} & \multicolumn{1}{c}{\bf CER(\%)} &\multicolumn{1}{c}{\bf 5-scale SMOS}
\\ \hline
&Ground Truth     & 4.52 $\pm$ 0.05& 0.7 & 3.91 $\pm$ 0.07\\
&Mel + HiFi-GAN (\cite{kong2020hifi})& 4.28 $\pm$ 0.08& 0.75 & 3.86 $\pm$ 0.08\\
LibriTTS&Guided-TTS 2 & 4.20 $\pm$ 0.08 & 0.84 & 3.70 $\pm$ 0.09\\
Dataset&Guided-TTS 2 (zero-shot) & 4.25 $\pm$ 0.09 & 0.8 & 3.51 $\pm$ 0.10\\
&YourTTS (\cite{2021arXiv211202418C}) & 4.02 $\pm$ 0.10 & 2.38 & 3.30 $\pm$ 0.10\\
&Meta-StyleSpeech (\cite{pmlr-v139-min21b})& 3.98 $\pm$ 0.11 & 1.52 & 3.42 $\pm$ 0.09\\
\hline 
&Ground Truth     & 4.45 $\pm$ 0.05& 2.4 & 3.71 $\pm$ 0.07\\
&Mel + HiFi-GAN (\cite{kong2020hifi})& 4.21 $\pm$ 0.08& 2.81 & 3.72 $\pm$ 0.07\\
VCTK&Guided-TTS 2 & 4.11 $\pm$ 0.09 & 1.49 & 3.57 $\pm$ 0.10\\
Dataset&Guided-TTS 2 (zero-shot) & 4.23 $\pm$ 0.09 & 0.81 & 3.39 $\pm$ 0.09\\
&YourTTS (\cite{2021arXiv211202418C}) & 3.94 $\pm$ 0.10 & 2.36 & 3.19 $\pm$ 0.09\\
&Meta-StyleSpeech (\cite{pmlr-v139-min21b})& 3.65 $\pm$ 0.13 & 1.84 & 3.26 $\pm$ 0.10\\
\bottomrule
\end{tabular}
\vskip -0.5in
\end{center}
\end{table}

The results of each model for LibriTTS and VCTK are presented in Table \ref{mos_multi}. As can be seen from the results, Guided-TTS 2 (zero-shot) outperforms other zero-shot baselines in SMOS. This indicates that Guided-TTS 2 can more robustly generate speech similar to the reference for various speakers than other adaptive TTS models with zero-shot adaptation setting. Moreover, we find that adaptation performance improves with fine-tuning in terms of speaker similarity. We also confirm through MOS and CER that Guided-TTS 2 is superior to adaptive TTS models in sound quality and pronunciation accuracy, whether fine-tuning or not. Samples are available on the demo page.
\subsection{Analysis}
\textbf{Effect of Fine-tuning} Comparing the Guided-TTS 2 (zero-shot) and Guided-TTS 2 results in Table \ref{mos_LJSpeech} and \ref{mos_multi}, we observe that fine-tuning improves speaker similarity (SMOS) while losing pronunciation accuracy. In other words, as the model is fine-tuned, the sample gets closer to the reference speech when it comes to prosody and timbre while getting further away from the given text. Since the primary purpose of adaptive TTS is to generate a speech that is close to a given reference speaker, we focus on improving the similarity while minimizing the sacrifice of pronunciation accuracy. We compare the CER and SECS according to the number of iterations for fine-tuning in Section \ref{app::analysis}.

\textbf{Effects of Speaker Classifier-free Gradient Scale}
We also show the performance of our model according to the speaker classifier-free gradient scale $\gamma_S$ for LibriTTS. For the fine-tuning setting and the zero-shot adaptation setting, we measure CER and SECS for multiple gradient scales. We check the effect of gradient scale on pronunciation accuracy through CER and on speaker similarity through SECS between embeddings extracted from the speaker encoder. We explore the speaker gradient scale $\gamma_S$ within $[0.0, 1.0, 2.0, 3.0, 4.0, 5.0, 6.0]$.

\begin{figure*}[t]
    \centering
    \setlength{\tabcolsep}{10.0pt}
    \begin{tabular}{cc}
        \includegraphics[width=0.40\textwidth]{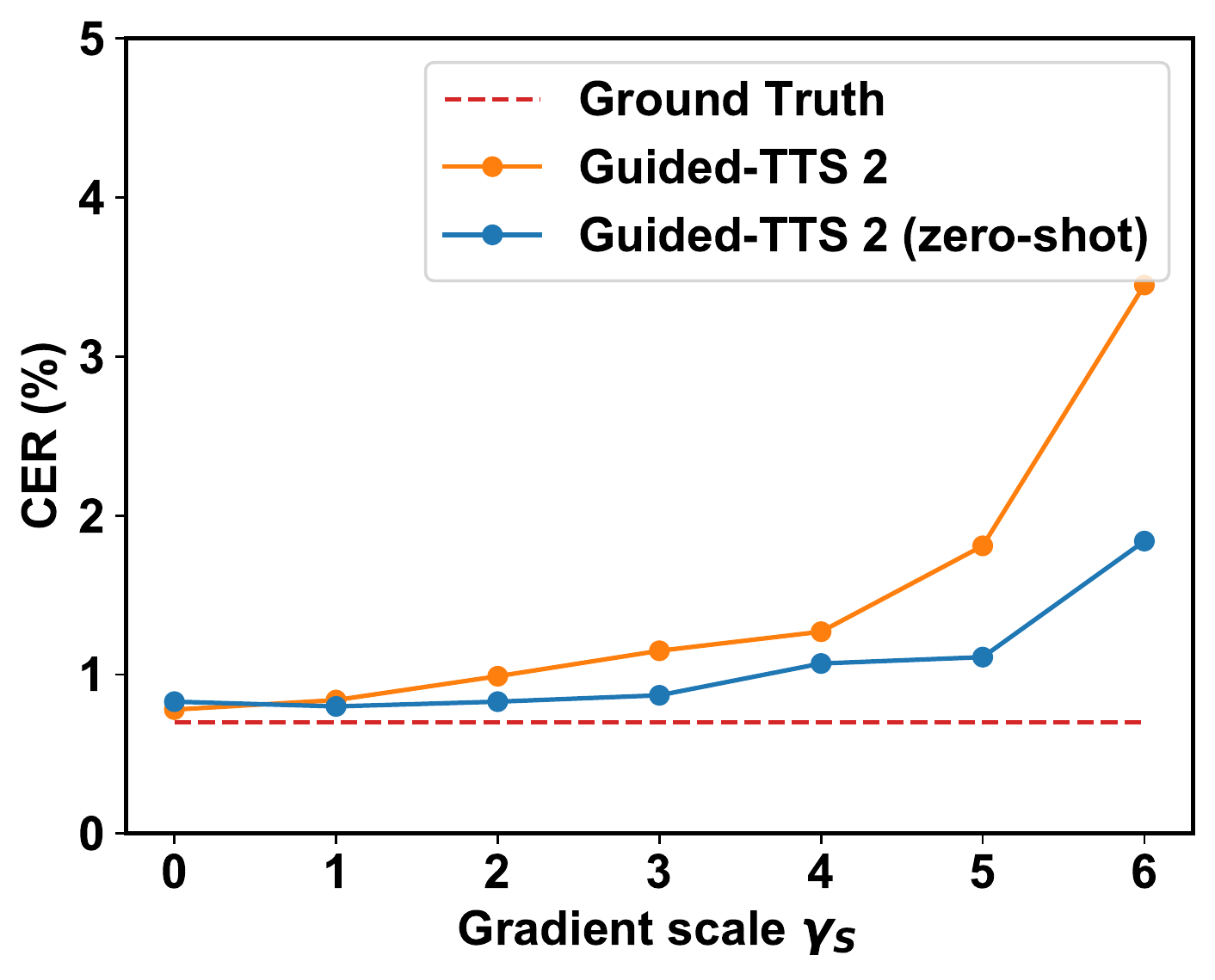} &
        \includegraphics[width=0.42\textwidth]{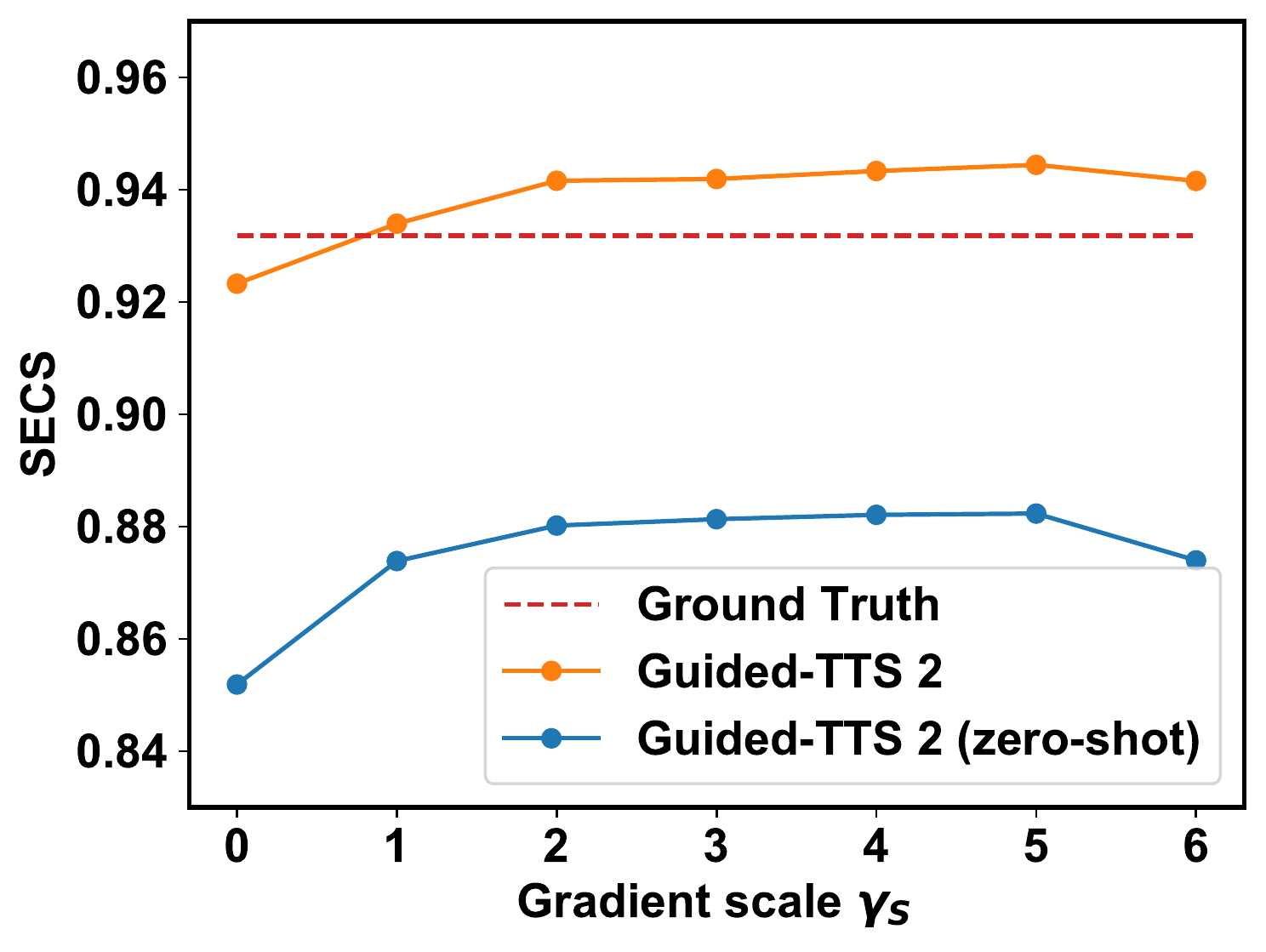} 
    \end{tabular}
    \caption{CER and SECS of Guided-TTS 2 according to speaker gradient scales. We refer to the fine-tuning setting of our model as Guided-TTS 2 and the zero-shot setting as Guided-TTS 2 (zero-shot).}
    \label{fig:speaker gradient scale}
    \vskip -0.2in 
\end{figure*}
Fig. \ref{fig:speaker gradient scale} presents the performance of each model for various gradient scales. As shown in Fig. \ref{fig:speaker gradient scale}, as the gradient scale increases, the pronunciation accuracy decreases, while the speaker similarity increases slightly. In particular, the pronunciation accuracy of our model with fine-tuning is sensitive to the gradient scale. We find that if the speaker gradient scale is set to $\gamma_S \in [1,3]$, Guided-TTS 2 can generate a high-quality sample with the target speaker's voice and accurate pronunciation. For convenience, we set the default gradient scale $\gamma_S$ to 1. 
\subsection{Text-to-Speech with Real-world Data}
To show the performance of Guided-TTS 2 in a real-world scenario, we adapt Guided-TTS 2 to various speakers using real-world audio clips as reference audio. We extract a 10-second audio clip of the desired speaker from Youtube and use it as reference data. With Guided-TTS 2, we can construct a high-quality TTS model using a wide range of voices such as Gollum in \textit{"The Lord of the Rings"}. We highly recommend readers to listen to the various samples on the demo page.
\section{Conclusion}
\label{conclusion}
In this work, we present Guided-TTS 2, a high-quality adaptive TTS model with only 10 seconds of the target speaker's unlabeled reference audio. As Guided-TTS 2 does not require any transcript to train the speaker-conditional diffusion model, we can leverage a large amount of untranscribed data. Using the classifier-free guidance and fine-tuning techniques for the diffusion-based model, Guided-TTS 2 can generate samples comparable to single-speaker TTS models with a short reference speech. We showed that Guided-TTS 2 is on par with the single-speaker TTS models for LJSpeech, and outperforms the adaptive TTS baselines for various datasets regarding sound quality and speaker similarity. We also showed that Guided-TTS 2 can generate samples robustly even with real-world clips such as Youtube clips. We believe that Guided-TTS 2 reduces the burden for high-quality TTS by reducing the amount of the target speaker's data.

Despite the advantages of Guided-TTS 2, some limitations are present. Guided-TTS 2 lacks parameter efficiency for each speaker, as all the parameters of speaker-conditional DDPM are fine-tuned for adaptation. As \namecite{chen2021adaspeech} propose, research on fine-tuning only the minimal amount of parameters without degrading adaptation performance can be considered a future work. In addition, as Guided-TTS 2 uses a pre-trained universal vocoder for raw wave generation, the sample quality of Guided-TTS 2 is bounded by the performance of the universal vocoder. When it comes to adaptation to unseen speakers, the sample quality of the universal vocoder is also degraded with respect to the recording environments. In order not to depend on the performance of the external neural network, research on end-to-end adaptive TTS will also be a promising future work.

\bibliography{main}

\begin{thebibliography}{50}
\providecommand{\natexlab}[1]{#1}
\providecommand{\url}[1]{\texttt{#1}}
\expandafter\ifx\csname urlstyle\endcsname\relax
  \providecommand{\doi}[1]{doi: #1}\else
  \providecommand{\doi}{doi: \begingroup \urlstyle{rm}\Url}\fi

\bibitem[Anderson(1982)]{Anderson1982-ny}
Brian D~O Anderson.
\newblock Reverse-time diffusion equation models.
\newblock \emph{Stochastic Process. Appl.}, 12\penalty0 (3):\penalty0 313--326,
  May 1982.

\bibitem[Bernard(2021)]{phonemizer20}
Mathieu Bernard.
\newblock Phonemizer.
\newblock \url{https://github.com/bootphon/phonemizer}, 2021.

\bibitem[Casanova et~al.(2021)Casanova, Shulby, Gölge, Müller, de~Oliveira,
  {Candido Jr.}, da~Silva~Soares, Aluisio, and Ponti]{casanova21b_interspeech}
Edresson Casanova, Christopher Shulby, Eren Gölge, Nicolas~Michael Müller,
  Frederico~Santos de~Oliveira, Arnaldo {Candido Jr.}, Anderson
  da~Silva~Soares, Sandra~Maria Aluisio, and Moacir~Antonelli Ponti.
\newblock {SC-GlowTTS: An Efficient Zero-Shot Multi-Speaker Text-To-Speech
  Model}.
\newblock In \emph{Proc. Interspeech 2021}, pages 3645--3649, 2021.
\newblock \doi{10.21437/Interspeech.2021-1774}.

\bibitem[{Casanova} et~al.(2021){Casanova}, {Weber}, {Shulby}, {Junior},
  {G{\"o}lge}, and {Antonelli Ponti}]{2021arXiv211202418C}
Edresson {Casanova}, Julian {Weber}, Christopher {Shulby}, Arnaldo~Candido
  {Junior}, Eren {G{\"o}lge}, and Moacir {Antonelli Ponti}.
\newblock {YourTTS: Towards Zero-Shot Multi-Speaker TTS and Zero-Shot Voice
  Conversion for everyone}.
\newblock \emph{arXiv e-prints}, art. arXiv:2112.02418, December 2021.

\bibitem[Chen et~al.(2021{\natexlab{a}})Chen, Tan, Li, Liu, Qin, sheng zhao,
  and Liu]{chen2021adaspeech}
Mingjian Chen, Xu~Tan, Bohan Li, Yanqing Liu, Tao Qin, sheng zhao, and Tie-Yan
  Liu.
\newblock Adaspeech: Adaptive text to speech for custom voice.
\newblock In \emph{International Conference on Learning Representations},
  2021{\natexlab{a}}.
\newblock URL \url{https://openreview.net/forum?id=Drynvt7gg4L}.

\bibitem[Chen et~al.(2021{\natexlab{b}})Chen, Zhang, Chun), Weiss, Norouzi,
  Dehak, and Chan]{wavegrad2}
Nanxin Chen, Yu~Zhang, Heiga Zen~(Byungha Chun), Ron~J. Weiss, Mohammad
  Norouzi, Najim Dehak, and William Chan.
\newblock Wavegrad 2: Iterative refinement for text-to-speech synthesis.
\newblock 2021{\natexlab{b}}.
\newblock URL \url{https://arxiv.org/abs/2106.09660}.

\bibitem[Chen et~al.(2019)Chen, Assael, Shillingford, Budden, Reed, Zen, Wang,
  Cobo, Trask, Laurie, Gulcehre, van~den Oord, Vinyals, and
  de~Freitas]{chen2018sample}
Yutian Chen, Yannis Assael, Brendan Shillingford, David Budden, Scott Reed,
  Heiga Zen, Quan Wang, Luis~C. Cobo, Andrew Trask, Ben Laurie, Caglar
  Gulcehre, Aäron van~den Oord, Oriol Vinyals, and Nando de~Freitas.
\newblock Sample efficient adaptive text-to-speech.
\newblock In \emph{International Conference on Learning Representations}, 2019.
\newblock URL \url{https://openreview.net/forum?id=rkzjUoAcFX}.

\bibitem[Chung et~al.(2018)Chung, Nagrani, and Zisserman]{Voxceleb2}
J.~S. Chung, A.~Nagrani, and A.~Zisserman.
\newblock Voxceleb2: Deep speaker recognition.
\newblock In \emph{INTERSPEECH}, 2018.

\bibitem[Cooper et~al.(2020)Cooper, Lai, Yasuda, Fang, Wang, Chen, and
  Yamagishi]{cooper2020zero}
Erica Cooper, Cheng-I Lai, Yusuke Yasuda, Fuming Fang, Xin Wang, Nanxin Chen,
  and Junichi Yamagishi.
\newblock Zero-shot multi-speaker text-to-speech with state-of-the-art neural
  speaker embeddings.
\newblock In \emph{ICASSP 2020-2020 IEEE International Conference on Acoustics,
  Speech and Signal Processing (ICASSP)}, pages 6184--6188. IEEE, 2020.

\bibitem[Dhariwal and Nichol(2021)]{dhariwal2021diffusion}
Prafulla Dhariwal and Alexander~Quinn Nichol.
\newblock Diffusion models beat {GAN}s on image synthesis.
\newblock In A.~Beygelzimer, Y.~Dauphin, P.~Liang, and J.~Wortman Vaughan,
  editors, \emph{Advances in Neural Information Processing Systems}, 2021.
\newblock URL \url{https://openreview.net/forum?id=AAWuCvzaVt}.

\bibitem[Donahue et~al.(2019)Donahue, McAuley, and Puckette]{WaveGAN}
C.~Donahue, J.~McAuley, and M.~Puckette.
\newblock Adversarial audio synthesis.
\newblock \emph{International Conference on Learning Representations (ICLR)},
  2019.

\bibitem[Donahue et~al.(2021)Donahue, Dieleman, Binkowski, Elsen, and
  Simonyan]{donahue2021endtoend}
Jeff Donahue, Sander Dieleman, Mikolaj Binkowski, Erich Elsen, and Karen
  Simonyan.
\newblock End-to-end {A}dversarial {T}ext-to-{S}peech.
\newblock In \emph{International Conference on Learning Representations}, 2021.
\newblock URL \url{https://openreview.net/forum?id=rsf1z-JSj87}.

\bibitem[gil Lee et~al.(2022)gil Lee, Kim, Shin, Tan, Liu, Meng, Qin, Chen,
  Yoon, and Liu]{lee2022priorgrad}
Sang gil Lee, Heeseung Kim, Chaehun Shin, Xu~Tan, Chang Liu, Qi~Meng, Tao Qin,
  Wei Chen, Sungroh Yoon, and Tie-Yan Liu.
\newblock Priorgrad: Improving conditional denoising diffusion models with
  data-dependent adaptive prior.
\newblock In \emph{International Conference on Learning Representations}, 2022.
\newblock URL \url{https://openreview.net/forum?id=_BNiN4IjC5}.

\bibitem[Ho and Salimans(2021)]{ho2021classifierfree}
Jonathan Ho and Tim Salimans.
\newblock Classifier-free diffusion guidance.
\newblock In \emph{NeurIPS 2021 Workshop on Deep Generative Models and
  Downstream Applications}, 2021.
\newblock URL \url{https://openreview.net/forum?id=qw8AKxfYbI}.

\bibitem[Ho et~al.(2020)Ho, Jain, and Abbeel]{DDPM}
Jonathan Ho, Ajay Jain, and Pieter Abbeel.
\newblock {Denoising Diffusion Probabilistic Models}.
\newblock In \emph{Advances in Neural Information Processing Systems 33: Annual
  Conference on Neural Information Processing Systems 2020, NeurIPS 2020,
  December 6-12, 2020, virtual}, volume~33. Curran Associates, Inc., 2020.

\bibitem[Ito(2017)]{ljspeech17}
Keith Ito.
\newblock The lj speech dataset.
\newblock \url{https://keithito.com/LJ-Speech-Dataset/}, 2017.

\bibitem[Jeong et~al.(2021)Jeong, Kim, Cheon, Choi, and
  Kim]{jeong21_interspeech}
Myeonghun Jeong, Hyeongju Kim, Sung~Jun Cheon, Byoung~Jin Choi, and Nam~Soo
  Kim.
\newblock {Diff-TTS: A Denoising Diffusion Model for Text-to-Speech}.
\newblock In \emph{Proc. Interspeech 2021}, pages 3605--3609, 2021.
\newblock \doi{10.21437/Interspeech.2021-469}.

\bibitem[Jia et~al.(2018)Jia, Zhang, Weiss, Wang, Shen, Ren, Chen, Nguyen,
  Pang, Lopez~Moreno, and Wu]{Verification}
Ye~Jia, Yu~Zhang, Ron Weiss, Quan Wang, Jonathan Shen, Fei Ren, zhifeng Chen,
  Patrick Nguyen, Ruoming Pang, Ignacio Lopez~Moreno, and Yonghui Wu.
\newblock Transfer learning from speaker verification to multispeaker
  text-to-speech synthesis.
\newblock In S.~Bengio, H.~Wallach, H.~Larochelle, K.~Grauman, N.~Cesa-Bianchi,
  and R.~Garnett, editors, \emph{Advances in Neural Information Processing
  Systems}, volume~31. Curran Associates, Inc., 2018.
\newblock URL
  \url{https://proceedings.neurips.cc/paper/2018/file/6832a7b24bc06775d02b7406880b93fc-Paper.pdf}.

\bibitem[{Kahn} et~al.(2020){Kahn}, {Rivière}, {Zheng}, {Kharitonov}, {Xu},
  {Mazaré}, {Karadayi}, {Liptchinsky}, {Collobert}, {Fuegen}, {Likhomanenko},
  {Synnaeve}, {Joulin}, {Mohamed}, and {Dupoux}]{librilight}
J.~{Kahn}, M.~{Rivière}, W.~{Zheng}, E.~{Kharitonov}, Q.~{Xu}, P.~E.
  {Mazaré}, J.~{Karadayi}, V.~{Liptchinsky}, R.~{Collobert}, C.~{Fuegen},
  T.~{Likhomanenko}, G.~{Synnaeve}, A.~{Joulin}, A.~{Mohamed}, and E.~{Dupoux}.
\newblock Libri-light: A benchmark for asr with limited or no supervision.
\newblock In \emph{ICASSP 2020 - 2020 IEEE International Conference on
  Acoustics, Speech and Signal Processing (ICASSP)}, pages 7669--7673, 2020.
\newblock \url{https://github.com/facebookresearch/libri-light}.

\bibitem[Kim et~al.(2021{\natexlab{a}})Kim, Kim, and
  Yoon]{DBLP:journals/corr/abs-2111-11755}
Heeseung Kim, Sungwon Kim, and Sungroh Yoon.
\newblock Guided-tts: Text-to-speech with untranscribed speech.
\newblock \emph{CoRR}, abs/2111.11755, 2021{\natexlab{a}}.
\newblock URL \url{https://arxiv.org/abs/2111.11755}.

\bibitem[Kim et~al.(2020)Kim, Kim, Kong, and Yoon]{kim2020glow}
Jaehyeon Kim, Sungwon Kim, Jungil Kong, and Sungroh Yoon.
\newblock Glow-{TTS}: A {G}enerative {F}low for {T}ext-to-{S}peech via
  {M}onotonic {A}lignment {S}earch.
\newblock \emph{Advances in Neural Information Processing Systems}, 33, 2020.

\bibitem[Kim et~al.(2021{\natexlab{b}})Kim, Kong, and Son]{kim2021conditional}
Jaehyeon Kim, Jungil Kong, and Juhee Son.
\newblock Conditional variational autoencoder with adversarial learning for
  end-to-end text-to-speech.
\newblock \emph{arXiv preprint arXiv:2106.06103}, 2021{\natexlab{b}}.

\bibitem[Kim et~al.(2019)Kim, Lee, Song, Kim, and Yoon]{kimflowavenet}
Sungwon Kim, Sang-Gil Lee, Jongyoon Song, Jaehyeon Kim, and Sungroh Yoon.
\newblock Flowavenet: A generative flow for raw audio.
\newblock In \emph{International Conference on Machine Learning}, pages
  3370--3378, 2019.

\bibitem[Kong et~al.(2020)Kong, Kim, and Bae]{kong2020hifi}
Jungil Kong, Jaehyeon Kim, and Jaekyoung Bae.
\newblock Hi{F}i-{GAN}: {G}enerative {A}dversarial networks for {E}fficient and
  {H}igh {F}idelity {S}peech {S}ynthesis.
\newblock \emph{Advances in Neural Information Processing Systems}, 33, 2020.

\bibitem[Kong et~al.(2021)Kong, Ping, Huang, Zhao, and Catanzaro]{DiffWave}
Zhifeng Kong, Wei Ping, Jiaji Huang, Kexin Zhao, and Bryan Catanzaro.
\newblock {DiffWave: A Versatile Diffusion Model for Audio Synthesis}.
\newblock In \emph{International Conference on Learning Representations}, 2021.

\bibitem[Kuchaiev et~al.(2019)Kuchaiev, Li, Nguyen, Hrinchuk, Leary, Ginsburg,
  Kriman, Beliaev, Lavrukhin, Cook, et~al.]{kuchaiev2019nemo}
Oleksii Kuchaiev, Jason Li, Huyen Nguyen, Oleksii Hrinchuk, Ryan Leary, Boris
  Ginsburg, Samuel Kriman, Stanislav Beliaev, Vitaly Lavrukhin, Jack Cook,
  et~al.
\newblock Nemo: a toolkit for building ai applications using neural modules.
\newblock \emph{arXiv preprint arXiv:1909.09577}, 2019.

\bibitem[Louppe(2019)]{resemblyzer}
Gilles Louppe.
\newblock Resemblyzer.
\newblock \url{https://github.com/resemble-ai/Resemblyzer}, 2019.

\bibitem[McAuliffe et~al.(2017)McAuliffe, Socolof, Mihuc, Wagner, and
  Sonderegger]{MFA}
Michael McAuliffe, Michaela Socolof, Sarah Mihuc, Michael Wagner, and Morgan
  Sonderegger.
\newblock Montreal forced aligner: Trainable text-speech alignment using kaldi.
\newblock In \emph{INTERSPEECH}, 2017.

\bibitem[Min et~al.(2021)Min, Lee, Yang, and Hwang]{pmlr-v139-min21b}
Dongchan Min, Dong~Bok Lee, Eunho Yang, and Sung~Ju Hwang.
\newblock Meta-stylespeech : Multi-speaker adaptive text-to-speech generation.
\newblock In Marina Meila and Tong Zhang, editors, \emph{Proceedings of the
  38th International Conference on Machine Learning}, volume 139 of
  \emph{Proceedings of Machine Learning Research}, pages 7748--7759. PMLR,
  18--24 Jul 2021.
\newblock URL \url{https://proceedings.mlr.press/v139/min21b.html}.

\bibitem[Nichol et~al.(2021)Nichol, Dhariwal, Ramesh, Shyam, Mishkin, McGrew,
  Sutskever, and Chen]{DBLP:journals/corr/abs-2112-10741}
Alex Nichol, Prafulla Dhariwal, Aditya Ramesh, Pranav Shyam, Pamela Mishkin,
  Bob McGrew, Ilya Sutskever, and Mark Chen.
\newblock {GLIDE:} towards photorealistic image generation and editing with
  text-guided diffusion models.
\newblock \emph{CoRR}, abs/2112.10741, 2021.
\newblock URL \url{https://arxiv.org/abs/2112.10741}.

\bibitem[Panayotov et~al.(2015)Panayotov, Chen, Povey, and Khudanpur]{7178964}
Vassil Panayotov, Guoguo Chen, Daniel Povey, and Sanjeev Khudanpur.
\newblock Librispeech: An asr corpus based on public domain audio books.
\newblock In \emph{2015 IEEE International Conference on Acoustics, Speech and
  Signal Processing (ICASSP)}, pages 5206--5210, 2015.
\newblock \doi{10.1109/ICASSP.2015.7178964}.

\bibitem[Popov et~al.(2021)Popov, Vovk, Gogoryan, Sadekova, and
  Kudinov]{Grad-TTS}
Vadim Popov, Ivan Vovk, Vladimir Gogoryan, Tasnima Sadekova, and Mikhail
  Kudinov.
\newblock {Grad-TTS: A Diffusion Probabilistic Model for Text-to-Speech}.
\newblock In \emph{Proceedings of the 38th International Conference on Machine
  Learning, {ICML} 2021, 18-24 July 2021, Virtual Event}, volume 139 of
  \emph{Proceedings of Machine Learning Research}, pages 8599--8608. {PMLR},
  2021.

\bibitem[Prenger et~al.(2019)Prenger, Valle, and
  Catanzaro]{prenger2019waveglow}
Ryan Prenger, Rafael Valle, and Bryan Catanzaro.
\newblock Waveglow: A flow-based generative network for speech synthesis.
\newblock In \emph{ICASSP 2019-2019 IEEE International Conference on Acoustics,
  Speech and Signal Processing (ICASSP)}, pages 3617--3621. IEEE, 2019.

\bibitem[Radford et~al.(2021)Radford, Kim, Hallacy, Ramesh, Goh, Agarwal,
  Sastry, Askell, Mishkin, Clark, Krueger, and Sutskever]{pmlr-v139-radford21a}
Alec Radford, Jong~Wook Kim, Chris Hallacy, Aditya Ramesh, Gabriel Goh,
  Sandhini Agarwal, Girish Sastry, Amanda Askell, Pamela Mishkin, Jack Clark,
  Gretchen Krueger, and Ilya Sutskever.
\newblock Learning transferable visual models from natural language
  supervision.
\newblock In Marina Meila and Tong Zhang, editors, \emph{Proceedings of the
  38th International Conference on Machine Learning}, volume 139 of
  \emph{Proceedings of Machine Learning Research}, pages 8748--8763. PMLR,
  18--24 Jul 2021.
\newblock URL \url{https://proceedings.mlr.press/v139/radford21a.html}.

\bibitem[Ramesh et~al.(2022)Ramesh, Dhariwal, Nichol, Chu, and
  Chen]{https://doi.org/10.48550/arxiv.2204.06125}
Aditya Ramesh, Prafulla Dhariwal, Alex Nichol, Casey Chu, and Mark Chen.
\newblock Hierarchical text-conditional image generation with clip latents,
  2022.
\newblock URL \url{https://arxiv.org/abs/2204.06125}.

\bibitem[Ren et~al.(2019)Ren, Ruan, Tan, Qin, Zhao, Zhao, and
  Liu]{ren2019fastspeech}
Yi~Ren, Yangjun Ruan, Xu~Tan, Tao Qin, Sheng Zhao, Zhou Zhao, and Tie-Yan Liu.
\newblock Fast{S}peech: {F}ast, {R}obust and {C}ontrollable {T}ext to {S}peech.
\newblock volume~32, pages 3171--3180, 2019.

\bibitem[Ren et~al.(2021)Ren, Hu, Tan, Qin, Zhao, Zhao, and
  Liu]{ren2021fastspeech}
Yi~Ren, Chenxu Hu, Xu~Tan, Tao Qin, Sheng Zhao, Zhou Zhao, and Tie-Yan Liu.
\newblock Fast{S}peech 2: {F}ast and {H}igh-{Q}uality {E}nd-to-{E}nd {T}ext to
  {S}peech.
\newblock In \emph{International Conference on Learning Representations}, 2021.
\newblock URL \url{https://openreview.net/forum?id=piLPYqxtWuA}.

\bibitem[Ronneberger et~al.(2015)Ronneberger, Fischer, and
  Brox]{ronneberger2015u}
Olaf Ronneberger, Philipp Fischer, and Thomas Brox.
\newblock U-net: Convolutional networks for biomedical image segmentation.
\newblock In \emph{International Conference on Medical image computing and
  computer-assisted intervention}, pages 234--241. Springer, 2015.

\bibitem[Shen et~al.(2018)Shen, Pang, Weiss, Schuster, Jaitly, Yang, Chen,
  Zhang, Wang, Skerrv-Ryan, et~al.]{shen2018natural}
Jonathan Shen, Ruoming Pang, Ron~J Weiss, Mike Schuster, Navdeep Jaitly,
  Zongheng Yang, Zhifeng Chen, Yu~Zhang, Yuxuan Wang, Rj~Skerrv-Ryan, et~al.
\newblock Natural tts synthesis by conditioning wavenet on mel spectrogram
  predictions.
\newblock In \emph{2018 IEEE International Conference on Acoustics, Speech and
  Signal Processing (ICASSP)}, pages 4779--4783. IEEE, 2018.

\bibitem[Sohl-Dickstein et~al.(2015)Sohl-Dickstein, Weiss, Maheswaranathan, and
  Ganguli]{pmlr-v37-sohl-dickstein15}
Jascha Sohl-Dickstein, Eric Weiss, Niru Maheswaranathan, and Surya Ganguli.
\newblock Deep unsupervised learning using nonequilibrium thermodynamics.
\newblock In Francis Bach and David Blei, editors, \emph{Proceedings of the
  32nd International Conference on Machine Learning}, volume~37 of
  \emph{Proceedings of Machine Learning Research}, pages 2256--2265, Lille,
  France, 07--09 Jul 2015. PMLR.
\newblock URL \url{https://proceedings.mlr.press/v37/sohl-dickstein15.html}.

\bibitem[Song and Ermon(2019)]{NEURIPS2019_3001ef25}
Yang Song and Stefano Ermon.
\newblock Generative modeling by estimating gradients of the data distribution.
\newblock In H.~Wallach, H.~Larochelle, A.~Beygelzimer, F.~d\textquotesingle
  Alch\'{e}-Buc, E.~Fox, and R.~Garnett, editors, \emph{Advances in Neural
  Information Processing Systems}, volume~32. Curran Associates, Inc., 2019.
\newblock URL
  \url{https://proceedings.neurips.cc/paper/2019/file/3001ef257407d5a371a96dcd947c7d93-Paper.pdf}.

\bibitem[Song and Ermon(2020)]{NEURIPS2020_92c3b916}
Yang Song and Stefano Ermon.
\newblock Improved techniques for training score-based generative models.
\newblock In H.~Larochelle, M.~Ranzato, R.~Hadsell, M.F. Balcan, and H.~Lin,
  editors, \emph{Advances in Neural Information Processing Systems}, volume~33,
  pages 12438--12448. Curran Associates, Inc., 2020.
\newblock URL
  \url{https://proceedings.neurips.cc/paper/2020/file/92c3b916311a5517d9290576e3ea37ad-Paper.pdf}.

\bibitem[Song et~al.(2021)Song, Sohl-Dickstein, Kingma, Kumar, Ermon, and
  Poole]{song2021scorebased}
Yang Song, Jascha Sohl-Dickstein, Diederik~P Kingma, Abhishek Kumar, Stefano
  Ermon, and Ben Poole.
\newblock Score-based generative modeling through stochastic differential
  equations.
\newblock In \emph{International Conference on Learning Representations}, 2021.
\newblock URL \url{https://openreview.net/forum?id=PxTIG12RRHS}.

\bibitem[van~den Oord et~al.(2016)van~den Oord, Dieleman, Zen, Simonyan,
  Vinyals, Graves, Kalchbrenner, Senior, and Kavukcuoglu]{van2016wavenet}
A{\"{a}}ron van~den Oord, Sander Dieleman, Heiga Zen, Karen Simonyan, Oriol
  Vinyals, Alex Graves, Nal Kalchbrenner, Andrew~W. Senior, and Koray
  Kavukcuoglu.
\newblock Wavenet: {A} generative model for raw audio.
\newblock \emph{arXiv preprint arXiv:1609.03499}, 2016.

\bibitem[Wan et~al.(2018)Wan, Wang, Papir, and Moreno]{GE2E}
Li~Wan, Quan Wang, Alan Papir, and Ignacio~Lopez Moreno.
\newblock Generalized end-to-end loss for speaker verification.
\newblock In \emph{2018 IEEE International Conference on Acoustics, Speech and
  Signal Processing (ICASSP)}, pages 4879--4883, 2018.
\newblock \doi{10.1109/ICASSP.2018.8462665}.

\bibitem[Wang et~al.(2017)Wang, Skerry-Ryan, Stanton, Wu, Weiss, Jaitly, Yang,
  Xiao, Chen, Bengio, Le, Agiomyrgiannakis, Clark, and
  Saurous]{wang17n_interspeech}
Yuxuan Wang, R.J. Skerry-Ryan, Daisy Stanton, Yonghui Wu, Ron~J. Weiss, Navdeep
  Jaitly, Zongheng Yang, Ying Xiao, Zhifeng Chen, Samy Bengio, Quoc Le, Yannis
  Agiomyrgiannakis, Rob Clark, and Rif~A. Saurous.
\newblock {Tacotron: Towards End-to-End Speech Synthesis}.
\newblock In \emph{Proc. Interspeech 2017}, pages 4006--4010, 2017.
\newblock \doi{10.21437/Interspeech.2017-1452}.

\bibitem[Wu et~al.(2022)Wu, Tan, Li, He, Zhao, Song, Qin, and
  Liu]{wu2022adaspeech}
Yihan Wu, Xu~Tan, Bohan Li, Lei He, Sheng Zhao, Ruihua Song, Tao Qin, and
  Tie-Yan Liu.
\newblock Adaspeech 4: Adaptive text to speech in zero-shot scenarios.
\newblock \emph{arXiv preprint arXiv:2204.00436}, 2022.

\bibitem[Yamagishi et~al.(2019)Yamagishi, Veaux, and
  MacDonald]{Yamagishi2019CSTRVC}
Junichi Yamagishi, Christophe Veaux, and Kirsten MacDonald.
\newblock Cstr vctk corpus: English multi-speaker corpus for cstr voice cloning
  toolkit (version 0.92).
\newblock 2019.

\bibitem[Yan et~al.(2021)Yan, Tan, Li, Qin, Zhao, Shen, and
  Liu]{yan2021adaspeech}
Yuzi Yan, Xu~Tan, Bohan Li, Tao Qin, Sheng Zhao, Yuan Shen, and Tie-Yan Liu.
\newblock Adaspeech 2: Adaptive text to speech with untranscribed data.
\newblock In \emph{ICASSP 2021-2021 IEEE International Conference on Acoustics,
  Speech and Signal Processing (ICASSP)}, pages 6613--6617. IEEE, 2021.

\bibitem[Zen et~al.(2019)Zen, Dang, Clark, Zhang, Weiss, Jia, Chen, and
  Wu]{zen19_interspeech}
Heiga Zen, Viet Dang, Rob Clark, Yu~Zhang, Ron~J. Weiss, Ye~Jia, Zhifeng Chen,
  and Yonghui Wu.
\newblock {LibriTTS: A Corpus Derived from LibriSpeech for Text-to-Speech}.
\newblock In \emph{Proc. Interspeech 2019}, pages 1526--1530, 2019.
\newblock \doi{10.21437/Interspeech.2019-2441}.

\end{thebibliography}
\bibliographystyle{plainnat}

\appendix
\newpage
\section{Details}
\subsection{Architectures and Training Details}\label{app::architectures}
In this section, we describe the training details and architecture of each model not covered in the main text. For all the modules except for the speaker-conditional diffusion model, we use the same structure from Guided-TTS \shortcite{DBLP:journals/corr/abs-2111-11755}, and also train them the same way. Hyperparameters and training details of each module are in Table \ref{hparams}.

\begin{table}[h]
\caption{Hyperparameters of Guided-TTS 2.}
\label{hparams}
\small
\centering
\begin{tabular}{l|l|c}
\toprule
& \multicolumn{1}{c|}{Hyperparameter} & \multicolumn{1}{c}{Guided-TTS 2} \\ \midrule\midrule

\multirow{8}{*}{Phoneme Classifier} 
&Batch Size & 64\\ 
&Training Iterations & 1.1M\\
&WaveNet Residual Channel Size           & 256   \\
&WaveNet Residual Blocks                       & 6       \\
&WaveNet Dilated Layers                 & 3   \\
&WaveNet Dilation Rate                 & 2   \\
&$\beta_0$ for the Beta Schedule & 0.05 \\
&$\beta_1$ for the Beta Schedule& 20 \\\midrule 
\multirow{2}{*}{Duration Predictor}
&Batch Size & 64\\
&Training Iterations & 330K\\\midrule 
\multirow{5}{*}{Speaker Encoder}
&Batch Size & 64\\ 
&Training Iterations & 300K\\ 
&LSTM Layers                       & 2       \\
&LSTM Channels                 & 768   \\
&Speaker Embedding Dimension                 & 256   \\
\bottomrule
\end{tabular}
\end{table}

\textbf{Phoneme Classifier}
The phoneme classifier, for which we use the WaveNet \shortcite{van2016wavenet} architecture, is trained to classify corrupted mel-spectrograms as phoneme labels frame-wise with cross-entropy loss. Since the length of the phoneme label and the mel-spectrogram should be the same for frame-wise classification, we use the alignment extracted from the Montreal Forced Aligner (MFA) \shortcite{MFA} to match the length. 

\textbf{Duration Predictor} 
The duration predictor is a module that predicts the duration of each phoneme. We train the duration predictor with the L2 loss between the log-scale of the ground truth duration extracted from MFA and the predicted duration. We use a duration predictor with the same architecture as Glow-TTS \shortcite{kim2020glow}.

\textbf{Speaker Encoder}
We utilize speaker embedding for training the speaker-conditional diffusion model, speaker-dependent phoneme classifier, and duration predictor. Our speaker encoder compresses speaker information of mel-spectrogram into speaker embedding $e_S$ and is trained with GE2E \shortcite{GE2E} loss on the speaker verification dataset.

\subsection{Hardware and Sampling Speed}
We perform all experiments and evaluations using a NVIDIA RTX 8000 with 48GB memory. To check the sampling speed of Guided-TTS 2, we measure the real-time factor (RTF) for generating 8-second-long speech. When we use the classifier-free guidance method, our speaker-conditional DDPM estimates both the conditional and unconditional scores simultaneously, which makes the synthesis speed slower. If we do not use the classifier-free guidance method, the RTF of Guided-TTS 2 is 0.526. With the classifier-free guidance, the RTF of Guided-TTS 2 is 0.802. Even though the primary goal of Guided-TTS 2 is a high-quality adaptive TTS, Guided-TTS 2 enables real-time synthesis. 

\subsection{Subjective Evaluation}
We evaluate the sample quality and the speaker similarity of each model by measuring MOS and SMOS via Amazon Mechanical Turk (AMT). For MOS and SMOS evaluation, we provide reference audio clips for 1, 3, and 5 points to listeners for each assessment, and we strongly recommend wearing headphones to evaluate. When measuring MOS, we provide listeners with the prompt \textit{"How natural (i.e., human-sounding) is this recording? Please focus on the audio quality and the naturalness of pronunciation."} and the ground truth transcript, and we ask them to evaluate the sample quality of the audio clips for the transcript on a scale of 1 to 5 points. For measuring SMOS, we provide listeners with the prompt \textit{"How similar are the speaker in the reference audio and the speakers in each recording? (Please focus on the tone of voice and speaking style rather than audio quality.)"} and reference audio from the target speaker and ask them to rate the speaker similarity between the reference audio and provided audio clips on a scale of 1 to 5. We spent a total of around $\$900$ for the subjective evaluations.

\section{Additional Results}
\subsection{Comparison on LJSpeech}\label{app::LJ_16kHz}
In Section \ref{results::single}, we show that our model achieves comparable performance to single-speaker TTS models, Grad-TTS \shortcite{Grad-TTS} and Guided-TTS. In order to show that the existing adaptive TTS baselines are far inferior to the single-speaker TTS baselines regarding sample quality and speaker similarity, we also compare our model and single-speaker TTS models with the existing adaptive baselines, YourTTS \shortcite{2021arXiv211202418C} and Meta-StyleSpeech \shortcite{pmlr-v139-min21b}, on LJSpeech. We measure MOS, SMOS, and CER to check sample quality, speaker similarity, and pronunciation accuracy for comparison. We choose reference audios and generate samples in the same way as in Section \ref{results::single}. Since adaptive TTS models generate 16kHz audio, all samples are downsampled to 16kHz for a fair comparison. 

\begin{table}[h]
\caption{Mean opinion score (MOS) and speaker similarity mean opinion score (SMOS) with 95$\%$ confidence intervals of single-speaker TTS models, adaptive TTS models, and Guided-TTS 2 for LJSpeech. The diffusion model of Guided-TTS 2 in this table is trained on LibriTTS \shortcite{zen19_interspeech} and Libri-Light \shortcite{librilight}.}
\label{mos_LJSpeech_16kHz}
\begin{center}
\begin{tabular}{lcccc}
\toprule
\multicolumn{1}{c}{\bf Method} &\multicolumn{1}{c}{\bf 5-scale MOS} & \multicolumn{1}{c}{\bf CER(\%)} &\multicolumn{1}{c}{\bf 5-scale SMOS}
\\ \hline 
Ground Truth     & 4.43 $\pm$ 0.05 & 0.64 & 3.89$\pm$0.06\\
Mel + HiFi-GAN (\cite{kong2020hifi})& 4.22 $\pm$ 0.08 & 0.86 & 3.85$\pm$0.06\\
Grad-TTS (\cite{Grad-TTS})   &  4.22 $\pm$ 0.08 & 0.98 & 3.84$\pm$0.07\\
Guided-TTS (\cite{DBLP:journals/corr/abs-2111-11755})&  4.18 $\pm$ 0.07 & 1.23 & 3.75$\pm$0.07\\
\hline
Guided-TTS 2  & 4.19 $\pm$ 0.08 & 1.12 & 3.77$\pm$0.07\\
Guided-TTS 2 (zero-shot)  & 4.14 $\pm$ 0.08 & 0.89 & 3.66$\pm$0.07\\
YourTTS (\cite{2021arXiv211202418C})   & 3.99 $\pm$ 0.10 & 2.76 & 3.43$\pm$0.09\\
Meta-StyleSpeech (\cite{pmlr-v139-min21b}) & 3.84 $\pm$ 0.11 & 1.52 & 3.45$\pm$0.09\\
\bottomrule
\end{tabular}
\end{center}
\end{table}
The performance of each model is in Table \ref{mos_LJSpeech_16kHz}. Each metric value of YourTTS and Meta-StyleSpeech indicates a significant performance gap between the existing adaptive baselines and single-speaker TTS models. On the other hand, Guided-TTS 2 outperforms adaptive TTS baselines and is comparable to single-speaker TTS models. These results demonstrate that only 10 seconds of reference audio is enough to construct a high-quality TTS model, which is on par with TTS models trained on a sufficient amount of data from the target speaker.

\subsection{Analysis for Fine-tuning}\label{app::analysis}
\textbf{Effect of Length of Reference Audio}
To analyze the effect of the length of the reference sample on pronunciation accuracy and speaker similarity, we measure CER and SECS with samples generated using reference audios of various lengths. As in the previous section, we conduct experiments on LJSpeech. For the case of reference audio shorter than 10 seconds, we cut audios of LJSpeech to 3 or 5 seconds. For the case of reference audio longer than 10 seconds, we concatenate several audios into 30 or 60 seconds-long samples and use them as reference samples.

\begin{table}[h]
\caption{CER and SECS of Guided-TTS 2 according to the length of reference audios.}
\label{amount_of_data}
\begin{center}
\begin{tabular}{lcccc}
\toprule
\multicolumn{1}{c}{\bf Method} & \multicolumn{1}{c}{\bf Length(seconds)} & \multicolumn{1}{c}{\bf CER(\%)}&\multicolumn{1}{c}{\bf SECS}
\\ 
\hline
  & 3 & 2.44 & 0.925 \\
  & 5 & 1.67 & 0.930 \\
Guided-TTS 2  & 10 & 1.12 & 0.929 \\
 & 30 & 0.98 & 0.932 \\
 & 60 & 1.14 & 0.931 \\
\bottomrule
\end{tabular}
\end{center}
\end{table}

The performance of each length model is in Table \ref{amount_of_data}. We find that pronunciation accuracy improves by using longer reference audio until a length of about 10 seconds. In addition, we observe that speaker similarity is maintained well even if the length of the reference sample used is about 3-second-long. These results indicate that the length of the reference audio is critical for accurate pronunciation. In order to build a high-quality TTS model with shorter reference audio, a method to preserve pronunciation capability is required, which can be a possible future research direction.

\textbf{Effect of Optimizer and the Number of Fine-tuning Iterations}
We check the effects of the number of fine-tuning iterations and statistics of the pre-trained optimizer on fine-tuning. For evaluation, we use the same reference samples and sentences of the LibriTTS dataset in Section \ref{results::adaptive} and generated samples in the same way as in Section \ref{results::adaptive} except for whether the optimizer is reused and the number of iterations of fine-training. For comparison, we measure CER and SECS.

\begin{table}[h]
\caption{CER and SECS of Guided-TTS 2 according to the number of fine-tuning iterations and whether the pre-trained optimizer is loaded.}
\label{mos_ft}
\begin{center}
\begin{tabular}{lcccc}
\toprule
\multicolumn{1}{c}{\bf Method} & \multicolumn{1}{c}{\bf Iterations} & \multicolumn{1}{c}{\bf Optimizer} & \multicolumn{1}{c}{\bf CER(\%)}&\multicolumn{1}{c}{\bf SECS}
\\ 
\hline
  & 0 & - & 0.8 & 0.873 \\
  & 50 & Initialize & 0.82 & 0.908 \\
Guided-TTS 2  & 200 & Initialize & 0.88 & 0.929 \\
  & 500 &Initialize &  0.84 & 0.937 \\
 & 2000 &Initialize &  1.49 & 0.945 \\
 \hline
Guided-TTS 2 & 500 & Load &  1.39 & 0.925 \\
\bottomrule
\end{tabular}
\end{center}
\end{table}
We observe two results in Table \ref{mos_ft}. First, as the number of fine-tuning iterations increases, speaker similarity gradually improves, while pronunciation accuracy deteriorates rapidly after a certain number of iterations. Second, using the statistics of the pre-trained optimizer for fine-tuning negatively affects both pronunciation accuracy and speaker similarity. Through this experiment, we find the optimal setting of Guided-TTS 2, which can generate samples as close to the reference audio as possible without pronunciation deterioration.

\section{Societal Impact}
Guided-TTS 2 has an advantage in significantly reducing  data required for high-quality adaptive TTS. In addition, Guided-TTS 2 can adapt to not only human voice but also non-human characters such as Gollum, which shows the possibility of extension to TTS for non-human characters in industries such as games and movies. On the other hand, 10-second untranscribed speech for the target speaker is easy to obtain through recording or YouTube clips for celebrities, and the contribution of Guided-TTS 2 that reduces the data required for high-quality adaptive TTS makes a lot of room for misuse. Guided-TTS 2 is likely to be misused as voice phishing for individuals or to have a fatal effect on the security system through voice. Given this potential misuse, we've decided not to release our code. Although we do not release the code, due to the adaptation ability of the diffusion-based model, we expect that the adaptive TTS technology is highly likely to be misused like Deepfake. We leave the research on anti-spoofing that distinguishes generated speech from real audio as future work, considering the potential for misuse of Guided-TTS 2.

\end{document}